\documentclass[aps,twocolumn,a4paper,floatfix,superscriptaddress,prl,longbibliography]{revtex4-1}
\usepackage[toc,page]{appendix}
\usepackage{braket}
\usepackage{epsfig,graphicx,times}
\usepackage{amstext}
\usepackage{amsmath}
\usepackage{amssymb}
\usepackage{mathrsfs}
\usepackage{dcolumn}
\usepackage{bm}
\usepackage[tight]{subfigure}
\usepackage[colorlinks,linkcolor=red,anchorcolor=blue,citecolor=blue,urlcolor=red]{hyperref}
\usepackage{color}
\usepackage{siunitx}
\usepackage{appendix}
\usepackage{placeins}
\usepackage{textcomp}
\usepackage{bbold}
\usepackage{float}
\usepackage{verbatim}

\begin{document}

\title{Nonreciprocal Optomechanical Entanglement against Backscattering Losses}

\author{Ya-Feng Jiao}
\affiliation{Key Laboratory of Low-Dimensional Quantum Structures and Quantum Control of Ministry of Education, Department of Physics and Synergetic Innovation Center for Quantum Effects and Applications, Hunan Normal University, Changsha 410081, China}

\author{Sheng-Dian Zhang}
\affiliation{Key Laboratory of Low-Dimensional Quantum Structures and Quantum Control of Ministry of Education, Department of Physics and Synergetic Innovation Center for Quantum Effects and Applications, Hunan Normal University, Changsha 410081, China}

\author{Yan-Lei Zhang}
\affiliation{CAS Key Laboratory of Quantum Information, University of Science and Technology of China, Hefei 230026, China}
\affiliation{CAS Center For Excellence in Quantum Information and Quantum Physics, University of Science and Technology of China, Hefei, Anhui 230026, China}

\author{Adam Miranowicz}
\affiliation{Faculty of Physics, Adam Mickiewicz University, 61-614 Pozna\'{n}, Poland}

\author{Le-Man Kuang}\email{lmkuang@hunnu.edu.cn}
\affiliation{Key Laboratory of Low-Dimensional Quantum Structures and Quantum Control of Ministry of Education, Department of Physics and Synergetic Innovation Center for Quantum Effects and Applications, Hunan Normal University, Changsha 410081, China}

\author{Hui Jing}\email{jinghui73@gmail.com}
\affiliation{Key Laboratory of Low-Dimensional Quantum Structures and Quantum Control of Ministry of Education, Department of Physics and Synergetic Innovation Center for Quantum Effects and Applications, Hunan Normal University, Changsha 410081, China}

\date{\today}

\begin{abstract}
We propose how to achieve nonreciprocal quantum entanglement of light and motion and reveal its counterintuitive robustness against random losses. We find that by splitting the counterpropagating lights of a spinning resonator via the Sagnac effect, photons and phonons can be entangled strongly in a chosen direction but fully uncorrelated in the other. This makes it possible both to realize quantum nonreciprocity even in the absence of any classical nonreciprocity and also to achieve significant entanglement revival against backscattering losses in practical devices. Our work provides a way to protect and engineer quantum resources by utilizing diverse nonreciprocal devices, for building noise-tolerant quantum processors, realizing chiral networks, and backaction-immune quantum sensors.
\end{abstract}

\maketitle

Nonreciprocal physics has witnessed rapid advances in recent years, with unique applications ranging from backaction-immune signal transfer or processing, chiral networking, and invisible sensing~\cite{Shoji2014STAM}. By breaking the Lorentz reciprocity, one-way flow of classical information, i.e., mean photon numbers, has been realized by using atoms~\cite{Ramezani2018PRL,Zhang2018NP}, solid devices~\cite{Dong2015NC,Kim2015NP,Xu2020prapp,Manipatruni2009PRL,Shen2016NP,Bernier2017NC,Mercier2019prapp,Fang2017NP,Hafezi2012OE}, and synthetic materials~\cite{Wang2013PRL,Yang2019Science,Maayani2018Nature,Sounas2017NP,Peng2014NP,Peng2016PNAS,Zhong2019OL}. Likewise, quantum optical diode or one-way flow of quantum information can also be achieved. In fact, nonreciprocal control of single photons and their quantum fluctuations have been demonstrated, such as single-photon diodes~\cite{Xia2018prl,Dong2019arxiv} or circulators~\cite{Scheucher2016Science}, and one-way photon blockade~\cite{Huang2018PRL,Yang2019arxiv}, providing key tools for chiral quantum engineering~\cite{Lodahl2017Nature,Ballestero2015prb,Gangaraj2017pra,Hu2019NP}. However, up to now, the possibilities of switching a single nonreciprocal device between classical and quantum regimes, as well as protecting quantum entanglement with nonreciprocal devices, have not yet been revealed.

\begin{figure*}[htbp]
\centering
\includegraphics[width=0.92\textwidth]{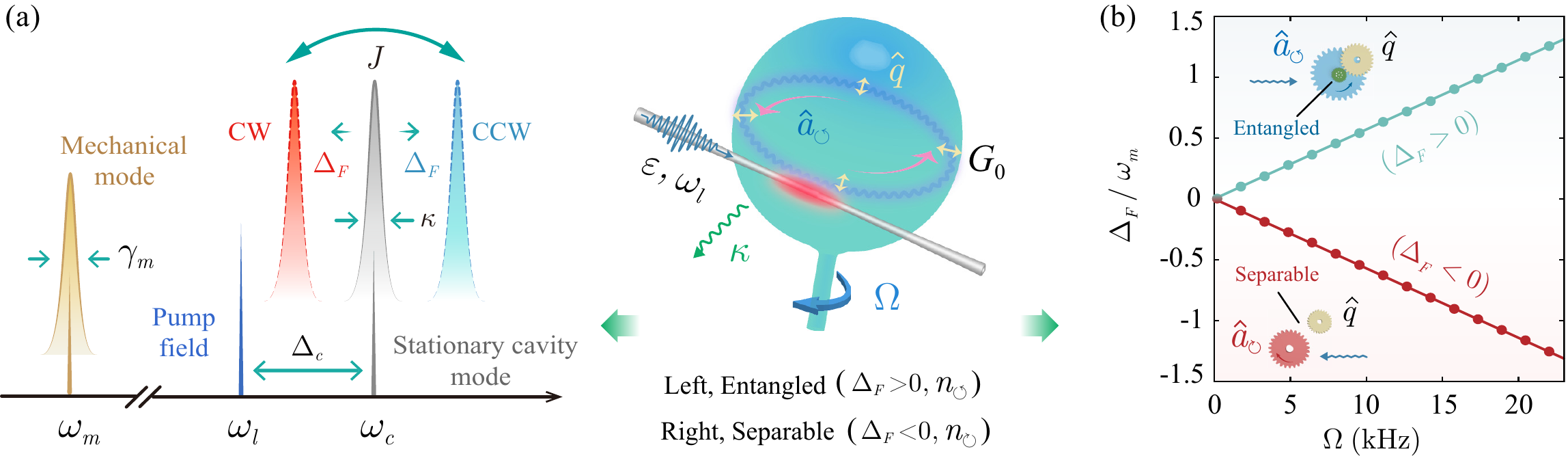}
\caption{\label{Fig.1}Nonreciprocal optomechanical entanglement in a spinning resonator. (a) Frequency spectrum of the spinning COM system. By fixing the CW rotation of the resonator, we have $ \Delta_{F}>0 $ ($ \Delta_{F}<0 $) for the case by driving the CCW (CW) mode. Besides, the resonator can support a radiation-pressure-induced mechanical radial breathing mode. The CW and CCW modes are coupled via backscattering with the strength $ J $. (b) The Sagnac-Fizeau shift $ \Delta_{F} $ versus the angular velocity $ \Omega $. Increasing the angular velocity results in a linear opposite frequency shift for the counterpropagating modes. For the same input light, due to the opposite frequency shift for the counterpropagating modes, COM entanglement can appear unidirectionally. See the text for more details.}
\end{figure*}

Here we propose how to achieve nonreciprocal quantum entanglement in cavity optomechanics (COM), revealing its unique properties which are otherwise unattainable in conventional devices. COM devices featuring coherent light-motion coupling~\cite{Aspelmeyer2014RMP,Verhagen2012Nature} have been widely used for quantum control of massive objects~\cite{Hong2017Science,Lecocq2015prx,Wollman2015Science,Pirkkalainen2015prl,Stannigel2010prl,Mirhosseini2020arxiv}, particularly COM entanglement~\cite{Vitali2007PRL,Huang2009NJP,Genes2008PRA,Riedinger2016Nature,Palomaki2013Science,Barzanjeh2019Nature,Chen2020NC,Korppi2018Nature,Riedinger2018Nature} or COM sensors~\cite{Massel2012NC,McClelland2011LPR,Qvarfort2018NC}. Very recently, quantum correlations at room temperature were observed even between light and \SI{40}{\kilo\gram} mirrors~\cite{Yu2020Nature}. Here we show that COM entanglement can be manipulated in a highly asymmetric way and the resulting nonreciprocal entanglement has the counterintuitive ability to preserve its optimal quality in a chosen direction against losses. This gives a new way to engineer quantum resources by utilizing diverse nonreciprocal devices, without the need of any topological or dissipative structure. In a broader view, our findings shed new light on the marriage of nonreciprocal physics and quantum technology, which can benefit such a wide range of applications as noise-tolerant quantum processing~\cite{Stannigel2012PRL,Horodecki2009RMP}, chiral quantum networking~\cite{Lodahl2017Nature,Ballestero2015prb,Gangaraj2017pra,Hu2019NP,Kimble2008Nature}, and backaction-immune quantum sensing~\cite{Ma2017NP,Fleury2015NC,Yang2015AM}.

As shown in Fig.\,\ref{Fig.1}, we consider a spinning COM resonator evanescently coupled with a tapered fiber. In a recent experiment~\cite{Maayani2018Nature}, nonreciprocal propagation of light with 99.6 $ \% $ isolation was demonstrated by using such a spinning device. The optical paths of counterpropagating lights in the resonator are different due to the rotation, resulting in an irreversible refractive index for the clockwise (CW) and counterclockwise (CCW) modes~\cite{Maayani2018Nature}, i.e., $ n_{\circlearrowleft,\circlearrowright}\!=\!n[1\pm nR\Omega(n^{-2}-1)/c] $, where $ n $ is the refractive index of the material, $ \Omega $ is the angular velocity of the resonator with radius $ R $, and $ c $ is the speed of light in the vacuum. Correspondingly, the resonance frequencies of the counterpropagating modes experience an opposite Sagnac-Fizeau shift; i.e., $ \omega_{c}\rightarrow\omega_{c}+\Delta_{F} $, with~\cite{Malykin2000PU}
\begin{equation}\small
\Delta_{F}=\pm\Omega\dfrac{nR\omega_{c}}{c}\left(1-\dfrac{1}
{n^{2}}-\dfrac{\lambda}{n}\dfrac{\mathrm{d}n}{\mathrm{d}\lambda}\right), \label{eq1}
\end{equation}
where $ \omega_{c} $ is the resonance frequency for a stationary resonator, and $ \lambda $ is the light wavelength in vacuum. The dispersion term $ dn/d\lambda $, characterizing the relativistic origin of the Sagnac effect, is small in typical materials (up to $ \sim1\% $)~\cite{Maayani2018Nature,Malykin2000PU}. By spinning the resonator along the CW direction, we have $ \Delta_{F}>0 $ or $ \Delta_{F}<0 $ for the case with the driving laser on the left- or right-hand side, and the corresponding effective optical frequencies are $ \omega_{j}\!\equiv\!\omega_{c}\pm|\Delta_{F}| $ ($ j=\circlearrowleft,\circlearrowright $), respectively. In addition, the resonator can support a mechanical breathing mode with frequency $ \omega_{m} $. In a rotating frame with respect to $ \hat{H}_{0}=\hbar\omega_{l}(\hat{a}_{\circlearrowleft}^{\dagger}\hat{a}_{\circlearrowleft}+\hat{a}_{\circlearrowright}^{\dagger}\hat{a}_{\circlearrowright})  $, the Hamiltonian of this spinning COM system, with the driving laser on the left-hand side, is this:
\begin{align}
\nonumber
&\hat{H}=\hat{\cal{H}}_{c}+\dfrac{\hbar\omega_{m}}{2}(\hat{p}^{2}+\hat{q}^{2})
-\hbar G_{0}(\hat{a}_{\circlearrowright}^{\dagger}\hat{a}_{\circlearrowright}
+\hat{a}_{\circlearrowleft}^{\dagger}\hat{a}_{\circlearrowleft})\hat{q},\\
&\hat{\cal{H}}_{c}\!=\!\sum_{j=\circlearrowleft,\circlearrowright}\!\hbar\Delta_{j}\hat{a}_{j}^{\dagger}\hat{a}_{j}\!+\!
\hbar J(\hat{a}_{\circlearrowright}^{\dagger}\hat{a}_{\circlearrowleft}
+\hat{a}_{\circlearrowleft}^{\dagger}\hat{a}_{\circlearrowright})
+i\hbar\varepsilon(\hat{a}_{\circlearrowleft}^{\dagger}
\!-\!\hat{a}_{\circlearrowleft}), \label{eq2}
\end{align}
where $ \hat{a}_{j} $ ($ \hat{a}_{j}^{\dagger} $) is the optical annihilation (creation) operator, $ \Delta_{j}=\omega_j-\omega_{l} $, and $ \hat{q} $ ($ \hat{p} $) is the dimensionless mechanical displacement (momentum) operator. The frame rotating with driving frequency $ \omega_{l} $ is obtained by applying the unitary transformation $\hat{U}=\exp[i\hat{H}_{0}t/\hbar]$ (see, e.g., Ref.\,\cite{Aspelmeyer2014RMP}). The field amplitude of the driving laser is $|\varepsilon|=\sqrt{2\kappa P/\hbar\omega_{l}} $, where $ P $ and $ \kappa $ are the input laser power and the optical decay rate, respectively. $G_{0}=(\omega_{c}/R)\sqrt{\hbar/m\omega_{m}}$ denotes the single-photon COM coupling rate~\cite{Aspelmeyer2014RMP}, with $ m $ the mass of the resonator. Also, imperfections of devices, such as surface roughness or material defect, can cause optical backscattering, as described by the mode-coupling strength $ J $. In a recent experiment, by breaking the time-reversal symmetry with Brillouin devices, dynamical suppression of the backscattering was already observed~\cite{Kim2019optica}.

Quantum Langevin equations of this spinning COM system then read:
\begin{align}
\nonumber
&\dot{\hat{a}}_{\circlearrowleft}=-(i\Delta_{\circlearrowleft}+\kappa)\hat{a}_{\circlearrowleft}
-iJ\hat{a}_{\circlearrowright}+iG_{0}\hat{a}_{\circlearrowleft}\hat{q}+\varepsilon+\sqrt{2\kappa}\hat{a}_{\circlearrowleft}^{\textrm{in}},\nonumber\\
&\dot{\hat{a}}_{\circlearrowright}=-(i\Delta_{\circlearrowright}+\kappa)\hat{a}_{\circlearrowright}-iJ\hat{a}_{\circlearrowleft}+iG_{0}\hat{a}_{\circlearrowright}\hat{q}
+\sqrt{2\kappa}\hat{a}_{\circlearrowright}^{\textrm{in}},\nonumber\\
&\dot{\hat{q}}=\omega_{m}\hat{p}, \nonumber\\ &\dot{\hat{p}}=-\omega_{m}\hat{q}-\gamma_{m}\hat{p}+G_{0}(\hat{a}_{\circlearrowright}^{\dagger}\hat{a}_{\circlearrowright}
+\hat{a}_{\circlearrowleft}^{\dagger}\hat{a}_{\circlearrowleft})+\hat{\xi},
\label{eq3}
\end{align}
where $ \gamma_{m} $ is the mechanical damping rate, and $ \hat{a}_{j}^{\textrm{in}} $ ($ \hat{\xi} $) is the zero-mean input noise operator for the optical (mechanical) mode, characterized by the following correlation functions~\cite{Zoller2000book}:
\begin{align}
\nonumber
&\langle \hat{a}_{j}^{\textrm{in}}(t)\hat{a}_{j}^{\textrm{in},\dagger}(t')\rangle=\delta(t-t'),\\
&\langle\hat{\xi}(t)\hat{\xi}(t')\rangle\simeq\gamma_{m}(2n_{m}+1)\delta(t-t'),~\text{for}~\omega_{m}/\gamma_{m}\!\gg\!1,
\label{eq4}
\end{align}
where $ n_{m}\!=\![\exp(\hbar\omega_{m}/k_{\textit{B}}\mathrm{T})-1]^{-1} $ denotes the thermal phonon number, $ k_{\textit{B}} $ is the Boltzmann constant, and $ \mathrm{T} $ is the bath temperature. Under the condition of strong optical driving, we can linearize the dynamics by expanding each operator as a sum of its steady-state value and a small fluctuation around it, i.e., $ \hat{a}_{j}=\alpha_{j}+\delta\hat{a}_{j},~\hat{q}=q_{s}+\delta\hat{q},~\hat{p}=p_{s}+\delta\hat{p} $. By defining the vectors of quadrature fluctuations and input noises as $ u^{\textit{T}}(t)=(\delta\hat{X}_{\circlearrowleft}, \delta\hat{Y}_{\circlearrowleft}, \delta\hat{X}_{\circlearrowright}, \delta\hat{Y}_{\circlearrowright}, \delta\hat{q}, \delta\hat{p}) $, $ v^{\textit{T}}(t)=(\sqrt{2\kappa}\hat{X}_{\circlearrowleft}^{\textrm{in}},\sqrt{2\kappa}\hat{Y}_{\circlearrowleft}^{\textrm{in}},\sqrt{2\kappa}\hat{X}_{\circlearrowright}^{\textrm{in}},\sqrt{2\kappa}\hat{Y}_{\circlearrowright}^{\textrm{in}},0,\hat{\xi}) $, with the components:
\begin{align}
\nonumber
\delta\hat{X}_{j}&=\dfrac{1}{\sqrt{2}}(\delta\hat{a}_{j}^{\dagger}
+\delta\hat{a}_{j}),& \delta\hat{Y}_{j}&=\dfrac{i}{\sqrt{2}}
(\delta\hat{a}_{j}^{\dagger}-\delta\hat{a}_{j}),\\
\hat{X}_{j}^{\textrm{in}}&=\dfrac{1}{\sqrt{2}}
(\hat{a}_{j}^{\textrm{in}\dagger}+\hat{a}_{j}^{\textrm{in}}),
&\hat{Y}_{j}^{\textrm{in}}&=\dfrac{i}{\sqrt{2}}(\hat{a}_{j}^{\textrm{in}\dagger}
-\hat{a}_{j}^{\textrm{in}}),
\label{eq6}
\end{align}
we obtain a compact form of the linearized equations of quantum fluctuations:
$
\dot{u}(t)=Au(t)+v(t), $
where
\begin{equation}
A=\begin{pmatrix}
-\kappa & \tilde{\Delta}_{\circlearrowleft} & 0 & J & -G_{\circlearrowleft}^{y} & 0\\
-\tilde{\Delta}_{\circlearrowleft} & -\kappa & -J & 0 & G_{\circlearrowleft}^{x} & 0\\
0 & J & -\kappa & \tilde{\Delta}_{\circlearrowright} & -G_{\circlearrowright}^{y} & 0\\
-J & 0 & -\tilde{\Delta}_{\circlearrowright} & -\kappa & G_{\circlearrowright}^{x} & 0\\
0 & 0 & 0 & 0 & 0 & \omega_{m}\\
G_{\circlearrowleft}^{x} & G_{\circlearrowleft}^{y} & G_{\circlearrowright}^{x} & G_{\circlearrowright}^{y} & -\omega_{m} & -\gamma_{m}
\end{pmatrix},
\label{eq8}
\end{equation}
and $ \tilde{\Delta}_{j}=\Delta_{j}-G_{0}q_{s} $ is the effective optical detuning, $ G_{j}^{x} $ ($ G_{j}^{y} $) is the real (imaginary) part of the effective COM coupling rate
($
G_{j}\equiv\sqrt{2}G_{0}\alpha_{j}=G_{j}^{x}+iG_{j}^{y}$),
and the steady-state mean values of the dynamical variables are given by:
\begin{align}
\nonumber
&\alpha_{\circlearrowleft}=\dfrac{(i\tilde{\Delta}_{\circlearrowright}+\kappa)\varepsilon}{(i\tilde{\Delta}_{\circlearrowright}+\kappa)(i\tilde{\Delta}_{\circlearrowleft}+\kappa)+J^{2}},\\ \nonumber
&\alpha_{\circlearrowright}=\dfrac{-iJ}{(i\tilde{\Delta}_{\circlearrowright}+\kappa)}\alpha_{\circlearrowleft},\\
&q_{s}=\dfrac{G_{0}}{\omega_{m}}(|\alpha_{\circlearrowright}|^{2}+|\alpha_{\circlearrowleft}|^{2}),~p_{s}=0.
\label{eq10}
\end{align}

The solution of the linearized Langevin equations is given by $ u(t)=\mathcal{M}(t)u(0)+\int_{0}^{t}d\tau \mathcal{M}(\tau)v(t-\tau) $, where $ \mathcal{M}(t)=\exp(At) $. The system is stable and reaches its steady state when all real parts of the  eigenvalues of $ A $ are negative, as characterized by the Routh-Hurwitz criterion~\cite{DeJesus1987PRA,SM}. When the stability condition is fulfilled, we have $ \mathcal{M}(\infty)=0 $ in the steady state and $  $
\begin{align}
u_{i}(\infty)=\int_{0}^{\infty}d\tau\sum_{k}\mathcal{M}_{ik}(\tau)v_{k}(t-\tau).\label{eq11}
\end{align}
Due to the linearized dynamics and the Gaussian nature of the quantum noises, the steady state of the system, independently of any initial conditions, finally evolves into a tripartite zero-mean Gaussian state, which is fully characterized by a $ 6\times6 $ correlation matrix $ V $, with its components
\begin{align}
V_{kl}=\langle u_{k}(\infty)u_{l}(\infty)\!+\!u_{l}(\infty)u_{k}(\infty)\rangle/2. \label{eq12}
\end{align}
By substituting Eq.\,(\ref{eq11}) into Eq.\,(\ref{eq12}) and using the fact that the six components of $ v(t) $ are uncorrelated, the steady-state correlation matrix $ V $ is obtained as
\begin{align}
V=\int_{0}^{\infty}d\tau\mathcal{M}(\tau)D\mathcal{M}^{T}(\tau), \label{eq13}
\end{align}
where $ D\!=\!\textrm{Diag}\,[\kappa,\kappa,\kappa,\kappa,0,\gamma_{m}(2n_{m}\!+\!1)] $ is the diffusion matrix, defined through $ \langle v_{k}(\tau)v_{l}(\tau')\!+\!v_{l}(\tau')v_{k}(\tau)\rangle/2=D_{kl}\delta(\tau-\tau') $. Under the stability condition, the steady-state correlation matrix $ V $ fulfills the Lyapunov equation~\cite{Vitali2007PRL}:
\begin{align}
AV+VA^{\textit{T}}=-D. \label{eq14}
\end{align}
This linear Eq.\,(\ref{eq14}) allows us to find $ V $ for any values of the relevant parameters. To quantify the entanglement between the mechanical mode and the driven optical mode, we adopt the logarithmic negativity, $ E_{\mathcal{N}} $, as a bipartite entanglement measure for continuous variables~\cite{Adesso2004pra}
\begin{align}
E_{\mathcal{N}}=\max\,[0,-\ln (2\nu^{-})], \label{eq16}
\end{align}
where $ \nu^{-}\!=\!2^{-1/2}\{\Sigma(V^{'})-[\Sigma(V^{'})^{2}-4\det\!V^{'}]^{1/2}\}^{1/2} $ , with $ \Sigma(V^{'})\!=\!\det\mathcal{A}+\det\mathcal{B}-2\det\mathcal{C} $, is the smallest eigenvalue of the partial transpose of the reduced correlation matrix $ V^{'} $. {By tracing out the reflected mode in $ V $ and writing it in a $ 2\times2 $ block form,} we have
\begin{align}
V^{'}=\left(
\begin{matrix}
\mathcal{A}&\mathcal{C}\\
\mathcal{C}^{\textit{T}}&\mathcal{B}
\end{matrix}
\right).\label{eq15}
\end{align}
$ V^{'} $ preserves the Gaussian nature, and entanglement emerges in its corresponding subsystem if and only if $ \nu^{-}<1/2 $, which is equivalent to Simon's necessary and sufficient entanglement criterion (or the related Peres-Horodecki criterion) for certifying entanglement of two-mode system in Gaussian states~\cite{Simon2000PRL}.

\begin{figure}[t]
	\centering
	\includegraphics[width=0.45\textwidth,height=0.56\textheight]{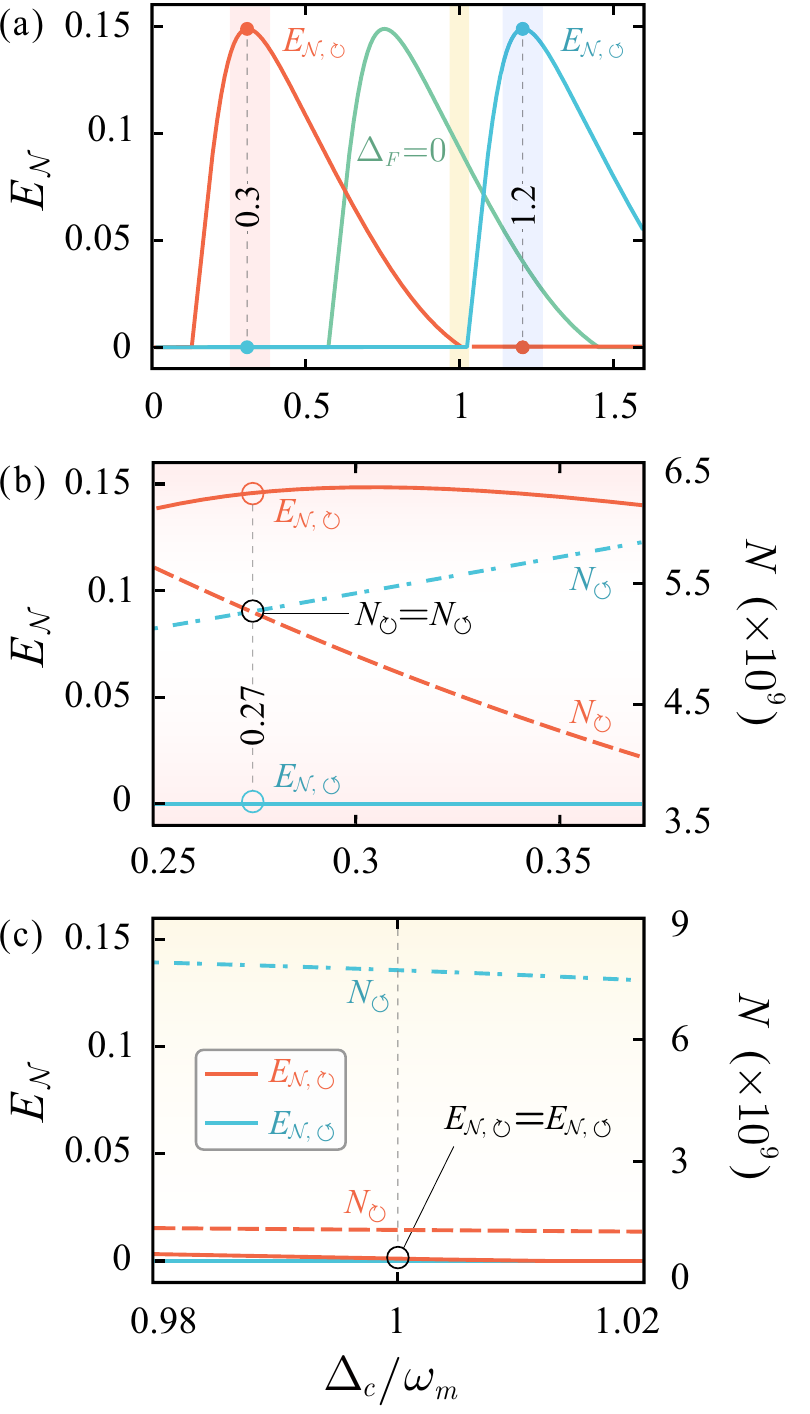}
	\caption{\label{Fig2}Nonreciprocal COM entanglement without backscattering. (a) $ E_{\mathcal{N}} $ versus $ \Delta_{c}/\omega_{m} $ for different input directions. For $ \Delta_F=0 $, COM entanglement exists around the resonance $ \Delta_{c}/\omega_{m}\simeq1 $. The spectral offset is due to the the COM-induced redshift of the cavity mode. For $ \Delta_F\neq0 $, the resonance conditions for the countercirculating modes are modified by the opposite Sagnac shifts, resulting in the peaks symmetrically shifted for the opposite input directions. (b)-(c) Tunable quantum nonreciprocity versus classical nonreciprocity. For $ \Delta_{c}/\omega_{m}\sim 0.27$, quantum nonreciprocity exists even when $N_\circlearrowleft=N_\circlearrowright$ (without classical nonreciprocity); in contrast, for $ \Delta_{c}/\omega_{m}\sim 1 $, classical nonreciprocity appears for $E_{\mathcal{N},\circlearrowleft}=E_{\mathcal{N},\circlearrowright}$. The parameters are chosen as $ \Omega = 8\,\textrm{kHz} $, $ J=0 $, and $ P = 20\,\textrm{mW} $. }		
\end{figure}

\begin{figure*}[t]
\centering
\includegraphics[width=0.9\textwidth,height=0.42\textheight]{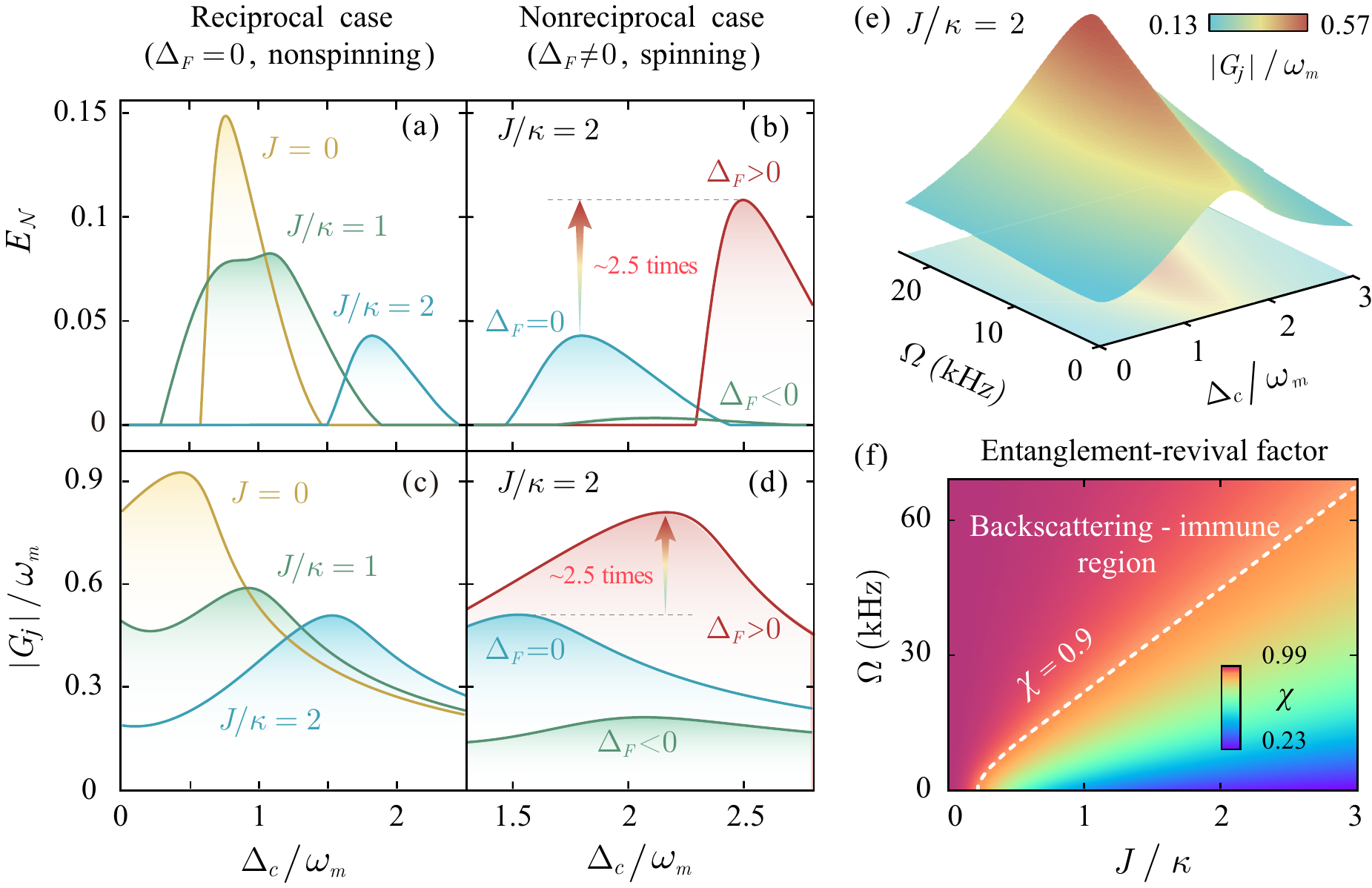}	
\caption{\label{Fig.3}Suppression of COM entanglement due to random-defect-induced backscattering, and its nonreciprocal revival resulting from the rotation-induced compensation. (a), (b) The logarithmic negativity $ E_{\mathcal{N}} $ and (c), (d) the effective COM coupling $ |G_{j}|/\omega_{m} $ are plotted as a function of the scaled optical detuning $ \Delta_{c}/\omega_{m} $. For $ \Omega=23\,\textrm{kHz} $, the value of $ E_{\mathcal{N}} $ is enhanced for $\sim 2.5$ times, reaching almost that as in an ideal device without backscattering. (e) The effective COM coupling $ |G_{j}|/\omega_{m} $ versus the optical detuning $ \Delta_{c}/\omega_{m} $ and the rotation speed $ \Omega $. (f) Density plot of the revival factor $ \chi $ as a function of the optical coupling strength $ J $ and the rotation speed $ \Omega $.}
\end{figure*}
In our calculations, for ensuring the stability of the system, we use the experimentally feasible values~\cite{Maayani2018Nature,Righini2011RNC}: $ n\!=\!1.48 $, $ m\!=\!10\,\textrm{ng} $, $ R\!=\!1.1\,\textrm{mm} $, $ \lambda\!=\!\SI{1.55}{\micro\metre} $, $ Q\!=\!\omega_{c}/\kappa\!=\!3.2\times10^{7} $, $ \omega_{m}\!=\!63\,\textrm{MHz} $, $ \gamma_{m}\!=\!5.2\,\textrm{kHz} $, $ \mathrm{T}\!=\!130\,\textrm{mK} $, and $ \Omega\!=\!8\,\textrm{kHz} $ or $ 23\,\textrm{kHz} $. We first consider the case without backscattering, i.e., $ J=0 $. In Fig.\,\ref{Fig2}, we plot the logarithmic negativity $ E_{\mathcal{N},j} $ and the intracavity photon number $ N_{j}\equiv|\alpha_{j}|^{2} $ of the driven mode, as a function of the detuning $ \Delta_{c}=\omega_{c}-\omega_{l} $. Here $ j $ denotes the driving direction. For a stationary resonator, $ E_{\mathcal{N},j} $ is independent on the driving direction, while for a spinning one, it becomes different by reversing the direction. For example, Figs.\,\ref{Fig2}(a) and \ref{Fig2}(b) show that, when the maximal COM entanglement is created by driving from one side, no entanglement occurs by driving it from the other side. The underlying physics can be understood as follows. In COM, the driving laser is scattered by the mechanical mode into the Stokes and anti-Stokes sidebands. When the cavity mode is resonant with one of the sidebands, COM correlations are created, as shown in experiments~\cite{Palomaki2013Science,Riedinger2016Nature}. Now, by spinning the resonator, nonreciprocity emerges for the created COM entanglement, which is fundamentally different from classical nonreciprocity of mean photon numbers. In fact, as shown in Figs.\,\ref{Fig2}(b) and \ref{Fig2}(c), nonreciprocal COM entanglement exists even without any classical nonreciprocity; in contrast, for $ \Delta_{c}/\omega_{m}\sim 1 $, significant classical nonreciprocity appears for $ E_{\mathcal{N},\circlearrowleft}\simeq E_{\mathcal{N},\circlearrowright} $. Hence it is possible to switch a single nonreciprocal device between classical and quantum regimes. We note that one-way quantum control of photon bunching and antibunching has been observed very recently~\cite{Scheucher2016Science,Yang2019arxiv}.

More importantly, quantum nonreciprocity provides a feasible way to protect devices against losses, which is reminiscent of that in topological systems or chiral environment~\cite{Gangaraj2017pra,Ballestero2015prb}. Figure\,\ref{Fig.3}(a) shows that, in conventional COM, $ E_{\mathcal{N}} $ decreases for $ J\neq 0 $, while for a spinning device, the maximum value of $ E_{\mathcal{N}} $ is significantly enhanced, approaching to that in an ideal device [see Fig.\,\ref{Fig.3}(b)]. To better understand this counterintuitive effect, we also plot the effective COM coupling of the driven mode $ |G_{j}| $ with respect to $ \Delta_{c} $ in Figs.\,\ref{Fig.3}(c)-\ref{Fig.3}(e). We see that the backscattering-induced reflection can be significantly suppressed in a spinning device. As a result, one can achieve nearly ideal COM entanglement, which can be clearly shown by defining a revival factor:
\begin{align}
\chi=\frac{\max\left[E_{\mathcal{N}}(\Omega\neq 0, J\neq 0)\right]}{\max\left[E_{\mathcal{N}}(\Omega=0, J=0)\right]}.
\end{align}
Figure\,\ref{Fig.3}(f) shows that the maximal factor can reach $99.1\%$; i.e., COM entanglement in such a nonreciprocal device is immune to random losses. This provides a new strategy to improve the performance of quantum devices by harnessing the power of nonreciprocity. Also we have confirmed that nonreciprocal entanglement can survive even when quantum COM entanglement is fully destroyed by thermal noises in a conventional reciprocal device (see the Supplemental Material for more details~\cite{SM}).

Finally, we remark that in experiments, COM entanglement can be detected by measuring the correlation matrix $ V $ under a proper readout choice via a filter~\cite{Genes2008PRA,Riedinger2016Nature,Palomaki2013Science}. The optical quadratures can be measured via a homodyne or heterodyne detection of the output~\cite{Palomaki2013Science,Barzanjeh2019Nature,Chen2020NC}, and the readout of mechanical ones requires a probe being resonant with the anti-Stokes sideband, mapping the mechanical motion to the output field~\cite{Palomaki2013Science,Korppi2018Nature}. With the same procedure, nonreciprocal features are observable also in the output field, including transmission rates~\cite{Lu2017PR} and COM correlations~\cite{Palomaki2013Science,Barzanjeh2019Nature,Chen2020NC,Riedinger2016Nature,Korppi2018Nature,Riedinger2018Nature}, detailed calculations of which will be given elsewhere.

In summary, we have shown how to achieve quantum nonreciprocal entanglement in a COM system, how to switch such a single nonreciprocal device between classical and quantum regimes, and how to keep the optimal entanglement in a chosen direction against losses. These findings, shedding light on the marriage of nonreciprocal physics and quantum engineering, open up the way to control quantum states by utilizing such diverse nonreciprocal devices as in optics, atomtronics~\cite{Dong2019arxiv,Scheucher2016Science,Yang2019arxiv}, electronics~\cite{Kurpiers2018Nature}, and in acoustics~\cite{Maznev2013WM}. In fact, quantum nonreciprocity is achievable in systems well beyond COM; for instance, one-way control of sub- or super-Poissonian correlations, as predicted in an optical system~\cite{Huang2018PRL}, was demonstrated very recently using cavity atoms~\cite{Yang2019arxiv}. In a broader view, nonreciprocal entanglement provides an unconventional tool for tasks that cannot be performed by classical one-way devices, such as building noise-tolerant quantum processors~\cite{Stannigel2012PRL} and achieving directional quantum sensing~\cite{Wollman2015Science,Pirkkalainen2015prl,Stannigel2010prl,Mirhosseini2020arxiv} or steering multipartite entanglement~\cite{Li2018prl,DiCarlo2010Nature,Qin2018PRL,Karg2020Science}.

We thank Chang-Ling Zou and Ran Huang for discussions. A. M. is supported by the Polish National Science Centre under the Maestro Grant No. DEC-2019/34/A/ST2/00081. L.-M. K. is supported by the National Natural Science Foundation of China (NSFC, Grant No. 11775075 and No. 11434011), and H. J. is supported by the NSFC (Grant No. 11774086 and No. 11935006). Y.-L. Z. is supported by the NSFC (Grants No. 11704370) and the China Postdoctoral Science Foundation (2019M652181).


%

\clearpage

\onecolumngrid

\setcounter{equation}{0} \setcounter{figure}{0}
\setcounter{table}{0} \setcounter{page}{1}\setcounter{secnumdepth}{3} \makeatletter
\renewcommand{\theequation}{S\arabic{section}.\arabic{equation}}
\renewcommand{\thefigure}{S\arabic{figure}}
\renewcommand{\bibnumfmt}[1]{[S#1]}
\renewcommand{\citenumfont}[1]{S#1}
\renewcommand\thesection{S\arabic{section}}
\makeatletter
\newcommand{\rmnum}[1]{\romannumeral #1}
\newcommand{\Rmnum}[1]{\expandafter\@slowromancap\romannumeral #1@}
\makeatother

\begin{center}
{\ \ }
\end{center}

\begin{center}
{\large \bf Supplementary Material for ``Nonreciprocal Optomechanical Entanglement \\ Against Backscattering Losses''}
\end{center}

\vspace{1mm}

\begin{center}
Ya-Feng Jiao,$^{1}$ Sheng-Dian Zhang,$^{1}$ Yan-Lei Zhang,$^{2,\,3}$ Adam Miranowicz,$^{4}$ Le-Man Kuang,$^{1,\,*}$ and Hui Jing$^{1,\,\dagger}$
\end{center}

\begin{minipage}[]{16cm}
\small{\it
\centering $^{1}$Key Laboratory of Low-Dimensional Quantum Structures and Quantum Control of Ministry of Education,  \\
\centering Department of Physics and Synergetic Innovation Center for Quantum Effects and Applications, \\
\centering Hunan Normal University, Changsha 410081, China \\
\centering $^{2}$CAS Key Laboratory of Quantum Information, University of Science and Technology of China, Hefei 230026, P. R. China \\
\centering $^{3}$CAS Center For Excellence in Quantum Information and Quantum Physics, University of Science and Technology of China, Hefei, Anhui 230026, P. R. China \\
\centering $^{4}$Faculty of Physics, Adam Mickiewicz University, 61-614 Pozna\'{n}, Poland \\
}
\end{minipage}

\vspace{8mm}
Here, we present the technical details on nonreciprocal optomechanical entanglement, including: (1) a summary of all parameter values used in our work; (2) experimental feasibility of our system, especially the conditions of stable coupling between the tapered fiber and the spinning resonator; (3) more discussions on mechanical or optical stability; (4) the influence of thermal effects and quality factors on nonreciprocal entanglement.
\vspace{8mm}

\tableofcontents

\clearpage

\section{System parameters}

Table~\ref{tab:parameters} shows the main symbols and parameters which have been used in this work.
\begin{table}[h]
\centering
\renewcommand\arraystretch{1.5}
\begin{tabular}{lllll}
\hline
\hline
{\bf Symbol} & {\bf Definition} & {\bf Name} & {\bf Value}
\\\hline
$\omega_{m}$ & & Mechanical resonance frequency &
\SI{63}{\mega\hertz} \\
$\gamma_{m}$ & & Mechanical linewidth &
\SI{5.2}{\kilo\hertz} \\
$Q_{m}$ & $\omega_{m}/\gamma_{m}$ & Mechanical quality factor &
\num{1.21e4}\\
$\mathrm{T}$ & & Mechanical bath temperature & \SI{130}{\milli\kelvin} \\
$n_m$ & $[\exp(\hbar\omega_{m}/k_{B}\mathrm{T})-1]^{-1}$
& Thermal phonon occupation & $269.4$ \\
$m$ & & Effective mass & \SI{10}{\nano\gram}\\
$x_{\mathrm{zp}}$ & $\sqrt{\hbar/(m\omega_{m})}$
& Zero point fluctuations & \SI{0.41}{\femto\metre}\\
$\omega_{c}$ & & Optical resonance frequency & \SI{1.22}{\peta\hertz} \\
$\lambda$ & & Laser wavelength & \SI{1.55}{\micro\metre} \\
$\kappa$ & & Cavity linewidth & \SI{38.0}{\mega\hertz}\\
$Q$ & $\omega_{c}/\kappa$ & Optical quality factor & \num{3.2e7}\\
$P$ & & Input laser power & \SI{20}{\milli\watt} or \SI{100}{\milli\watt}\\
$G_{0}$ & $ \omega_{c}x_{\mathrm{zp}}/R $ & Single-photon optomechanical
coupling rate & \SI{452.1}{\hertz}\\\hline
$R$ & & Sphere radius & \SI{1.1}{\milli\metre}\\
$r$ & & Fiber radius & \SI{544}{\nano\metre}\\
$\Omega$ & & Spinning frequency
& \SI{8}{\kilo\hertz} or \SI{23}{\kilo\hertz}\\
$\mathrm{E}$ & & Young modulus of silica
& \SI{75}{\giga\pascal}\\
$\Upsilon$ & & Elastic limit of silica & \SI{9}{\giga\pascal}\\
$\varepsilon_{0}$ & & Dielectric constant of air & $1$\\
$\varepsilon_{1}$ ($\varepsilon_{2}$) & & Dielectric
constant of silica & $3.9$\\
$n_{0}$ & & Refractive index of air & $1$\\
$n_{1}$ ($n_{2}$) & & Refractive index of silica & $1.48$
\\\hline\hline
\end{tabular}
\caption{Feasible parameters.
Unless specified otherwise, these parameters are applied to all evaluations in the text.}
\label{tab:parameters}
\end{table}

With these experimentally feasible parameters, we have confirmed that other quantum effects, such as two-mode squeezing or non-classicality, can also be manipulated in a highly asymmetric way in this optomechanical system, which may find applications in chiral quantum optics and technology~\cite{SLodahl2017Nature,SKimble2008Nature,SStannigel2012prl}.

\begin{figure}[t]
\centering
\includegraphics[width=0.9\textwidth]{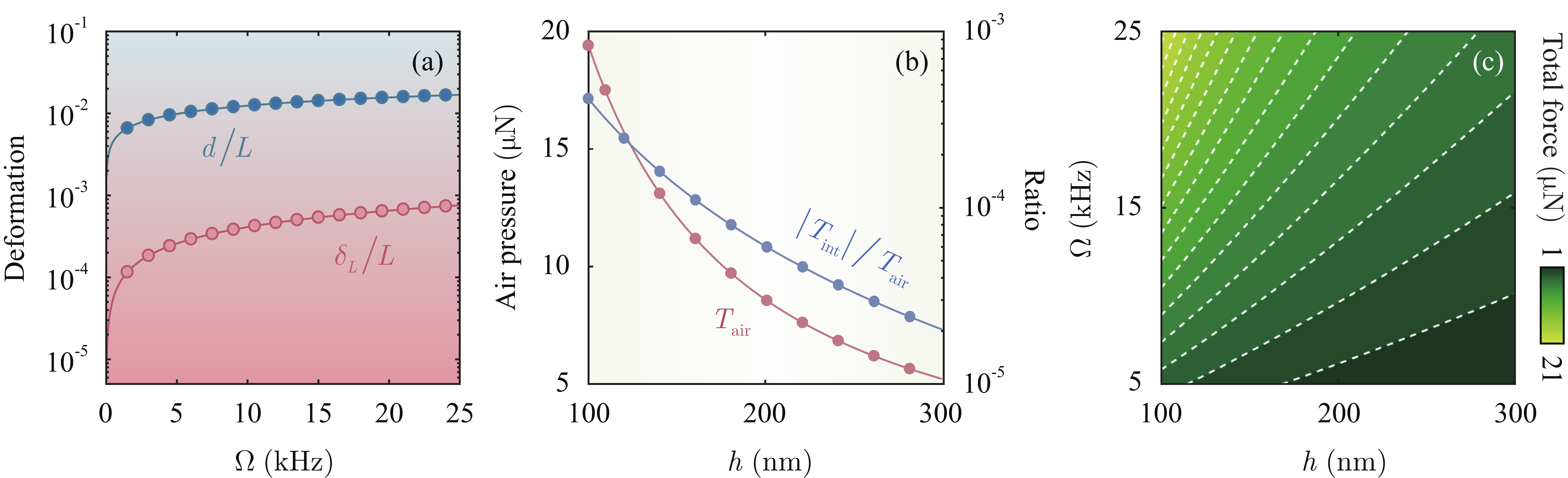}
\caption{\label{selfadjustment}Analysis of the ``self-adjustment'' behavior. (a) The strain $ \epsilon $ (red curve) and the displacement $ d $ (blue curve) as a function of the angular velocity $ \Omega $ for $ h = 250\,\textrm{nm} $. (b) The air pressure $ T_{\mathrm{air}} $ and the ratio of intermolecular forces $ |T_{\mathrm{int}}|/T_{\mathrm{air}} $ for varying taper-resonator separation $ h $ at $ \Omega=23\,\textrm{kHz} $. (c) Total force between the fiber and the sphere versus the angular velocity $ \Omega $ and the separation $ h $. The resulting force of air pressure and intermolecular forces has a minimal value of $T_{\mathrm{tot}} = \SI{1.135}{\micro\newton}$, which indicates the interactions of the fiber and the spinning sphere are always repulsive.}
\end{figure}

\section{Experimental feasibility}

\subsection{Self-adjustment process}

As shown in a very recent experiment~\cite{SMaayani2018Nature}, we consider a whispering-gallery-mode silica sphere, which is mounted on a turbine and spins along its axis with, e.g., the angular velocity $\Omega=6.6\,\textrm{kHz}$ for the sphere radius $ r=1.1\,\textrm{mm}$. Faster rotations have also been reported in experiments by using, e.g., levitated nanomechanical rotors~\cite{SReimann2018prl,SAhn2018prl}. Then, by positioning the spinning resonator near a single-mode telecommunication fiber, light can be coupled into or out of the resonator evanescently.

For such a spinning device, the aerodynamic process plays a key role in the stable fiber-resonator coupling. Specifically, the fast spinning resonator can drag air into the region between the tapered fiber and the resonator, thereby forming a lubrication layer of air in this region. Then the thin film of air, exerting pressure on the surface of the tapered fiber facing the resonator, can make the fiber fly above the resonator with the separation of a few nanometres. Therefore, if any perturbation causes the taper to rise higher than the stable-equilibrium height, it can float back to its original position, which is referred to as ``self-adjustment''. Hence the fiber will not touch or stick to the resonator even if it is pushed towards the spinning sphere~\cite{SMaayani2018Nature}. To see this clearly, we write the air pressure as $ \varDelta T_{\mathrm{air}} = (\rho\varDelta\theta)T_{\mathrm{air}}/L $, with $\rho$ ($\theta$) the radius (angle) of the winding shape of the deformation region. Then, the total air pressure $ T_{\mathrm{air}} $ on the taper can be estimated analytically as~\cite{SMaayani2018Nature}
\begin{align}
T_{\mathrm{air}}=6.19\mu R^{5/2}\Omega\int_{0}^{r}
\left(h-\sqrt{r^{2}-x^{2}}
+r\right)^{-3/2}\mathrm{d}x, \label{Tair}
\end{align}
where $\mu$ is the viscosity of air, $ R~(r) $ is the radius of the sphere (taper), and $ h=h_{0}+d $ represents the taper-resonator separation, with $h_{0}$ denoting the stationary gap between the fiber and the sphere. The local deformation can also lead to a tension on the infinitesimal cylinder of the fiber, which can be calculated by
\begin{align}
\varDelta T_{\mathrm{ela}} =2\mathcal{F}\sin\left(\varDelta\theta\middle/2\right)\approx \mathcal{F}\varDelta\theta, \label{Tela}
\end{align}
where $ \mathcal{F} $ is the elastic force on the taper, obeying $ \sigma=\mathcal{F}/(\pi r^{2})=\epsilon\mathrm{E} $. Here $ \sigma $ is the uniaxial stress, $\mathrm{E}$ is the Young modulus of silica, and $ \epsilon =\delta_{L}/L $ is the strain, where $ \delta_{L}=L'-L $ denotes the variation of the original length $ L $ of the deformation region. Furthermore, $ \delta_{L} $ can be straightforwardly derived via the following relations: $ L'=\rho\theta $, $ (L/2)^{2} + (\rho-d)^{2}=\rho^{2} $, and $ \sin(\theta/2)=L/(2\rho) $. Hence, in the case of equilibrium (i.e., $ \varDelta T_{\mathrm{air}}=\varDelta T_{\mathrm{ela}}$), $ T_{\mathrm{air}} $ can be given in another form:
\begin{align}
T_{\mathrm{air}}=2\pi r^{2}\mathrm{E}\left[\arcsin\left(\phi\right)-\phi\right]\approx\frac{\pi}{3} r^{2}\phi^{3}\mathrm{E}, \label{Tair2}
\end{align}
where $\left.\phi=4Ld\middle/\left.(L^{2} + 4d^{2}\right.)\right.$, and the approximation, $ \arcsin(\phi)=\phi +\phi^{3}/6+\cdots $, is made within the limit of $ |\phi|\ll 1 $, which physically requires a comparatively small distortion. As shown in Fig.\,\ref{selfadjustment}(a), we confirmed that these approximation conditions can be easily satisfied with experimentally accessible parameters~\cite{SMaayani2018Nature,SBellouard2005OE}: $\mathrm{E}=\SI{75}{\giga\pascal}$, $r=\SI{544}{\nano\metre}$,
and $L=\SI{3}{\micro\metre}$. In this case, the displacement $ d $ is given by
\begin{align}\label{displacement}
\left.d=\frac{L}{2}\left(\beta-\sqrt{\beta^{2}-1}\right),\right.
\end{align}
where $\left.\beta=\left[\pi r^{2}\mathrm{E} \middle/\left(3T_{\mathrm{air}}\right)\right]^{\left.1\middle/3\right.}\right.$. The strain of the taper thus can be reduced to $ \epsilon=\arcsin\left(\phi\right)/\phi-1\approx\phi^{2}/6 $, from which we find that the strain (i.e., the elastic force) is positively associated with the taper-resonator separation:
\begin{align}
\frac{\partial \mathcal{F}}{\partial h}
=\pi r^{2}\mathrm{E}\left(\frac{\partial \epsilon}
{\partial d}\right)
=\frac{16\pi r^{2}\mathrm{E}L^{2}d\left(L^{2}-4d^{2}\right)}
{3\left(L^{2}+4d^{2}\right)^{3}}>0. \label{strain}
\end{align}
Equation\,(\ref{strain}) clearly reveals that the elastic force becomes stronger when the air gap gets larger than the stable-equilibrium distance. Also, as shown in Fig.\,\ref{selfadjustment}(b), the air pressure on the taper is largely suppressed. As a result, the taper can be dragged back to its original position when any perturbation causes it away from the spinning resonator, leading to the self-adjustment behavior. The self-adjustment of the tapered fiber enables the critical coupling of light into or out of the resonator, by which the counter-propagating beams can experience an optical drag identical in size, but opposite in sign.

\subsection{Intermolecular forces}

The intermolecular forces, such as the Casimir and van der Waals forces, could also affect the stable fiber-resonator coupling. In our system, such kinds of intermolecular forces can be described as~\cite{SMaayani2018Nature}:
\begin{equation}\label{Casimir-forces}
T_{\mathrm{int}}=rR\left(-\frac{\mathbb{A}}{6\pi h^{3}}
+\frac{\mathbb{B}}{45\pi h^{9}}
-\frac{\pi^{2}c\hbar}{240h^{4}}\right),
\end{equation}
where the Hamaker constant $\mathbb{A}$ can be calculated by~\cite{SChen2009IEEE}:
\begin{equation}\label{Hamaker}
\mathbb{A}=\frac{3
\varepsilon_{-}^{\left(1\right)}
\varepsilon_{-}^{\left(2\right)}
k_{\mathrm{B}}\mathrm{T}}
{4\varepsilon_{+}^{\left(1\right)}
\varepsilon_{+}^{\left(2\right)}}
+\frac{\nu\left[n_{-}^{\left(1\right)}
n_{-}^{\left(2\right)}\right]^{2}
}{n_{+}^{\left(1\right)}
n_{+}^{\left(2\right)}
\left[n_{+}^{\left(1\right)}
+n_{+}^{\left(2\right)}\right]},
\end{equation}
with $\left.\nu=3\sqrt{2}\hbar\nu_{e}\middle/16\right.$, $\varepsilon_{\pm}^{\left(u\right)} = \varepsilon_{u}\pm\varepsilon_{0}$, $n_{\pm}^{\left(u\right)} = \sqrt{n_{u}^{2}\pm n_{0}^{2}}$, and $u=1$, $2$. Hereafter, we use $\varepsilon_{0}$ ($n_{0}$), $\varepsilon_{1}$ ($n_{1}$), and $\varepsilon_{2}$ ($n_{2}$) to represent the dielectric constant (the refractive index) of the air, the taper and the spinning resonator, respectively; $k_{\mathrm{B}}$ is the Boltzmann constant, $\mathrm{T}$ is the mechanical bath temperature, and $\nu_{e}=\SI{3}{\peta\hertz}$~\cite{SChen2009IEEE}. Note that for simplicity, we have replaced $n_{2}$ with $n$ in the main text. Moreover, the constant $\mathbb{B}$ is typically of the order of $\num{e-76}~\si{\joule.\metre^{6}}$ for interactions between condensed matter phases across the vacuum or air~\cite{SWu2002JT}. Taking intermolecular forces into account, the total force between the taper and the resonator becomes: $T_{\mathrm{tot}}=T_{\mathrm{air}}+T_{\mathrm{int}}$. Herein, we choose experimentally accessible parameters~\cite{SJoyce1968TSF}: $\varepsilon_{0}=1$, $\varepsilon_{1}=\varepsilon_{2}=3.9$, $n_{0}=1$, $n_{1}=n_{2}=1.48$, and $\mathrm{T}=\SI{130}{\milli\kelvin}$. As expected, the intermolecular forces are found to be negligible ($<0.1\%$), and the taper-resonator interactions remain repulsive [see Figs.~\ref{selfadjustment}(b) and \ref{selfadjustment}(c)]. Thus, the effects of the Casimir and van der Waals forces can be safely omitted on critical coupling. Other factors, such as lubricant compressibility, tapered-fiber stiffness, and the wrap angle of a fiber, may also affect critical coupling. However, these factors are confirmed to be negligible in experiments, which can be also safely ignored in our discussions~\cite{SMaayani2018Nature}.

\section{Stability conditions}

\subsection{Mechanical stability}

The realization of stable fiber-resonator coupling sets a limit to the angular velocity of spinning devices. Specifically, the requirement $\beta\geq 1$ in Eq.~\eqref{displacement} yields the first limit:
\begin{equation}
\Omega_{0}=\frac{\varrho\pi r^{2}\mathrm{E}}
{18.57\mu R^{\left.5\middle/2\right.}},
\end{equation}
where
\begin{equation}
\varrho=\left[\int_{0}^{r}\left(h-\sqrt{
	r^{2}-x^{2}}+r\right)^{\left.-3\middle/
	2\right.}\mathrm{d}x\right]^{-1}.
\end{equation}
Also, the tiny displacement should obey $d=h-h_{0}<h$, which gives the second limit:
\begin{equation}
\Omega_{1}=\frac{\varrho\pi r^{2}\Lambda\mathrm{E}}
{18.57\mu R^{\left.5\middle/2\right.}},
\end{equation}
where $\left.\Lambda=\left[4Lh\middle/\left(L^{2}+4h^{2}\right)\right]^{3}\right.$. Finally, the elastic limit of the tapered fiber provides the third stability condition ($\sigma=\Upsilon$):
\begin{equation}
\Omega_{2}=\frac{\varrho\pi r^{2}\Upsilon}
{3.095\mu R^{\left.5\middle/2\right.}}
\sqrt{\frac{6\Upsilon}{\mathrm{E}}},
\end{equation}
where $\Upsilon$ is typically \SI{9}{\giga\pascal} for silica devices~\cite{SSugiura1997JAP}. Thus, the mechanical limit to the angular velocity can be obtained as:
\begin{equation}
\Omega_{\max}=\min\left\{\Omega_{0},
\Omega_{1},\Omega_{2}\right\}.
\end{equation}
When operating at taper-resonator separations near \SI{250}{\nano\metre}, we have $\Omega_{0}=\SI{81.6}{\mega\hertz}$, $\Omega_{1}=\SI{2.8}{\mega\hertz}$, $\Omega_{2}=\SI{49.9}{\mega\hertz}$, and thus $\Omega_{\max}=\SI{2.8}{\mega\hertz}$. Therefore, it is reasonable to use $\Omega=\SI{8}{\kilo\hertz}$ or \SI{23}{\kilo\hertz} in the main text.

\begin{figure}[t]
    \centering
    \includegraphics[width=0.9\textwidth]{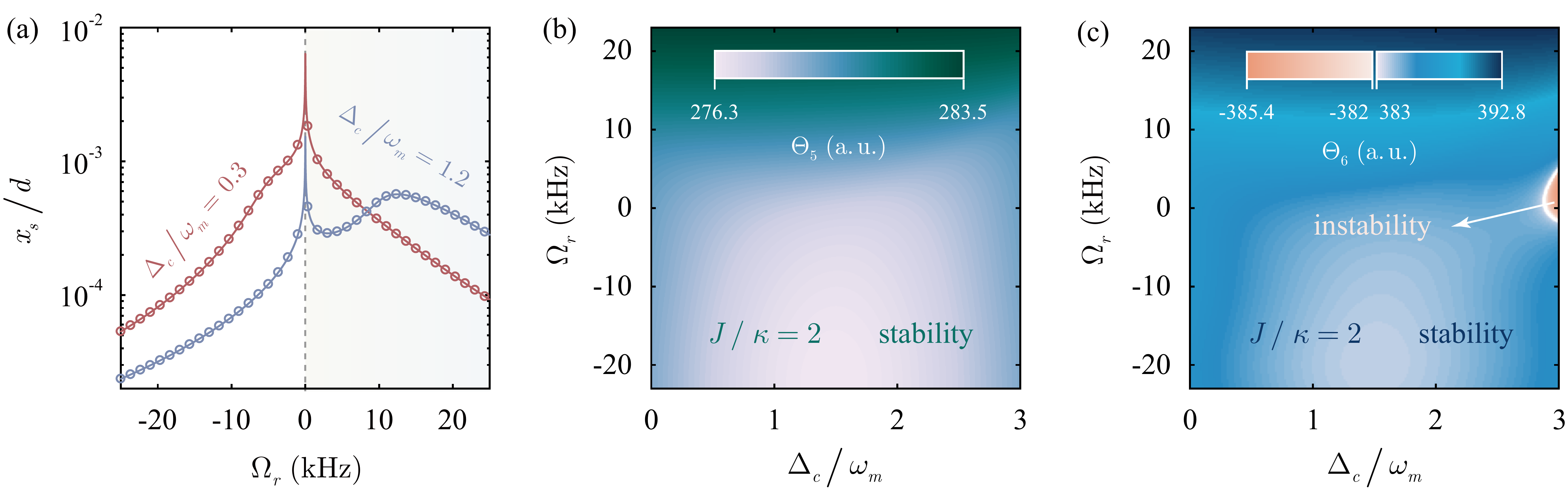}
    \caption{\label{stability}The mechanical and optical stability conditions. (a) The ratio of the mean mechanical displacement $ x_{s} $ to the air-induced displacement $ d $ as a function of the angular velocity $\Omega_{r}$. In the aerodynamic process, this ratio is extremely small ($ \sim 0.1\%$), meaning that radial breathing of the resonator is negligible compared to the air-induced displacement. (b-c) Stability functions $ \Theta_{5} $ and $ \Theta_{6} $ versus the angular velocity $ \Omega_{r} $ and the scaled optical detuning $ \Delta_{c}/\omega_{m} $ at $ J/\kappa=2 $ and $ P=20\,\textrm{mW}$. The white contour line in (c) is the boundary between the stability and instability regions, and a.u. denotes arbitrary units. The parameters are listed in Table~\ref{tab:parameters}.}
\end{figure}

Moreover, we also consider the influence of the radial breathing of the resonator on stable fiber-resonator coupling. In paticular, we compare the mean mechanical displacement $x_{s}$ with the air-induced displacement $d$, i.e.,
\begin{equation}
\left.\eta=x_{s}\middle/d=q_{s}x_{\mathrm{zp}}\middle/d,\right.
\end{equation}
where $x_{\mathrm{zp}}=\sqrt{\left.\hbar\middle/\left(m\omega_{m}\right)\right.}$ denotes the standard deviation of the zero-point motion of the mechanical mode. As shown in Fig.~\ref{stability}(a), the effect of the mechanical displacement is negligible indeed ($<1\%$), thereby the radial breathing of the resonator does not disturb the critical evanescent coupling. Note that, for simplicity, we have introduced $\Omega_{r}=\pm\Omega$.

\subsection{Optical stability}
\begin{figure}[t]
    \centering
    \includegraphics[width=0.9\textwidth]{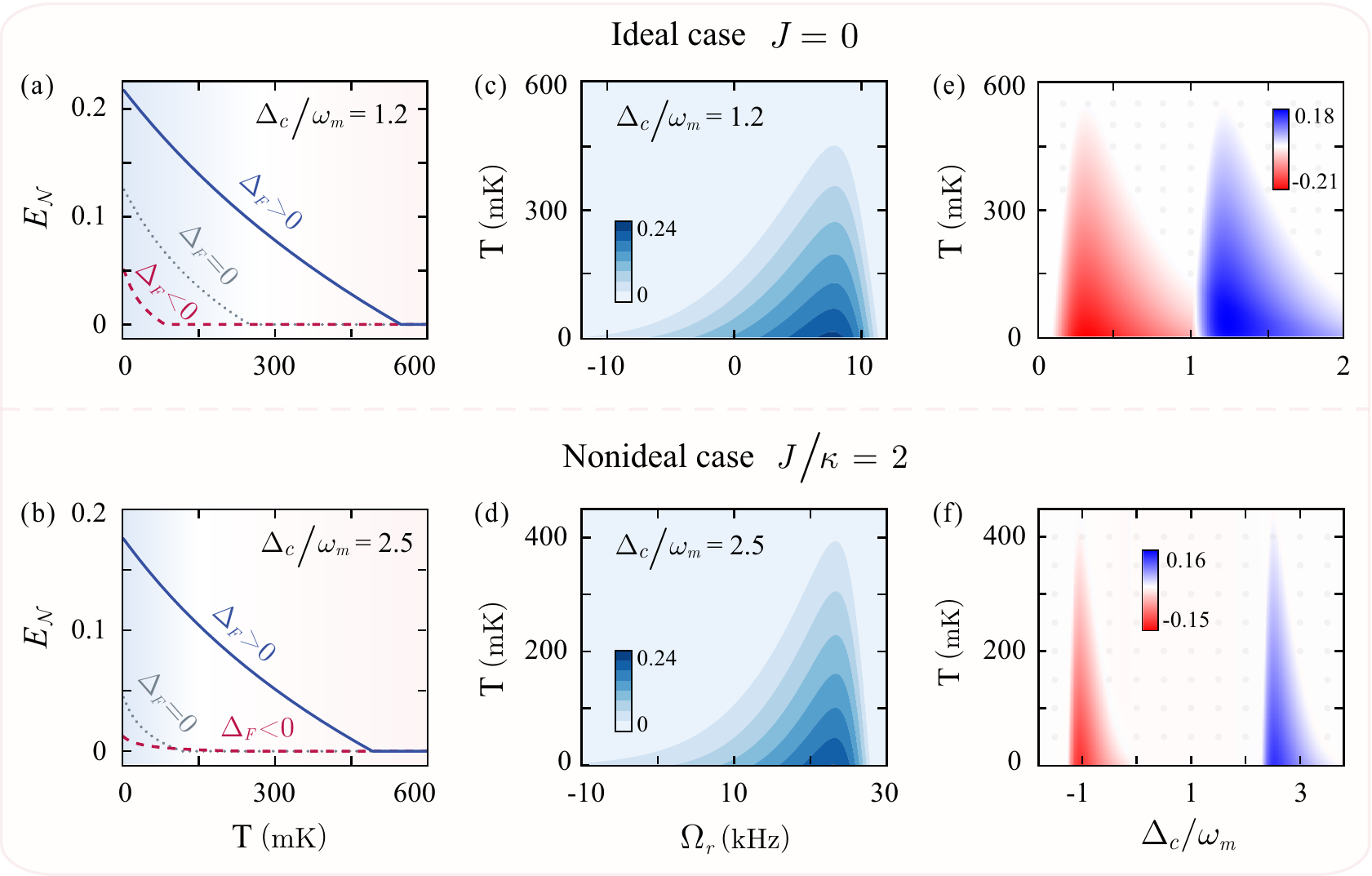}
    \caption{\label{thermaleffect}Thermal effect on nonreciprocal optomechanical entanglement. (a-b) The logarithmic negativity $ E_{\mathcal{N}} $ as a function of the environment temperature $ \textrm{T}$ for different driving directions. (c-d) Density plot of the logarithmic negativity $ E_{\mathcal{N}} $ versus the angular velocity $ \Omega_{r} $ and the environment temperature $ \textrm{T} $. (e-f) The COM entanglement difference with respect to different driving directions, $ \Delta E_{\mathcal{N}} $, versus the scaled optical detuning $ \Delta_{c}/\omega_{m} $ and the environment temperature $ \textrm{T} $. The rotation speed is chosen as $ \Omega=8\,\textrm{kHz} $ in (a) and (e), and $ \Omega=23\,\textrm{kHz} $ in (b) and (f). Other parameters are listed in Table~\ref{tab:parameters}.}		
\end{figure}
According to Routh-Hurwitz criterion~\cite{SDeJesus1987PRA}, the system is stable and reaches its steady state when all eigenvalues of the matrix $ A $ have negative real parts. Therefore, we start our analysis by determining the eigenvalues of the matrix $ A $, i.e., $ |A-\lambda I|=0 $, yielding the following characteristic equation:
\begin{align}
\lambda^{6}+a_{1}\lambda^{5}+a_{2}\lambda^{4}+a_{3}\lambda^{3}
+a_{4}\lambda^{2}+a_{5}\lambda+a_{6}=0,
\end{align}
where
\begin{align}
a_{1}&=4\kappa+\gamma_{m},~~~
a_{2}=\sigma_{0}+\omega_{m}^{2}
+4\kappa(\kappa+\gamma_{m}),\nonumber\\
a_{3}&=\sigma_{0}(2\kappa
+\gamma_{m})+4\kappa\left(
\omega_{m}^{2}
+\kappa\gamma_{m}\right),~~~
a_{4}=\sigma_{0}\sigma_{1}+\sigma_{2}
+4\kappa^{2}\omega_{m}^{2},\nonumber\\
a_{5}&=\gamma_{m}
\mu_{2}-2\kappa\mu_{0}\mu_{3}\left(\gamma_{m}\mu_{0}
+4\kappa\omega_{m}^{2}\right)+\kappa\mu_{1}
\left(\kappa\gamma_{m}+2\omega_{m}^{2}\right),\nonumber\\
a_{6}&=\omega_{m}\left(\sigma_{+}+\sigma_{-}+\sigma_{2}\omega_{m}
-\mu_{0}\mu_{4}\right)-\mu_{3}\tilde{\Delta}_{+}
\tilde{\Delta}_{-},
\end{align}
and
\begin{align}
\sigma_{0}&=2\mu_{0}+\mu_{1},~~~\sigma_{1}
=\kappa^{2}+2\kappa\gamma_{m}
+\omega_{m}^{2},~~~\sigma_{2}
=\mu_{0}^{2}+\kappa^{2}\mu_{1}+\mu_{2},\nonumber\\
\sigma_{2}&=\mu_{2}-\mu_{3}+\mu_{0}\left(J^{2}
-\kappa^{2}\right),~~~\sigma_{\pm}=
(\Delta_{\pm}J^{2}
-\Delta_{\mp}\kappa^{2})G_{\mp}^{2},
\nonumber\\
\mu_{0}&=J^{2}+\kappa^{2},~~~
\mu_{1}=\tilde{\Delta}_{+}^{2}+\tilde{\Delta}_{-}^{2},~~~
\mu_{2}=(\tilde{\Delta}_{+}\tilde{\Delta}_{-}-2J^{2})
\tilde{\Delta}_{+}\tilde{\Delta}_{-},\nonumber\\
\mu_{3}&=\omega_{m}(\tilde{\Delta}_{+}|
G_{\circlearrowleft}|^{2}+\tilde{\Delta}_{-}|
G_{\circlearrowright}|^{2}+\mu_{4}),~~~
\mu_{4}=2J(G_{\circlearrowright}^{x}
G_{\circlearrowleft}^{x}+G_{\circlearrowright}^{y}
G_{\circlearrowleft}^{y}).
\end{align}
Using the coefficients $ a_{k} $, we can form a set of $ k\times k$ matrices, $ \theta_{k} $, for $ k\leq 6 $, with their entries defined as:
\begin{align}
\theta_{ln}
= \begin{cases}
  0, & 2l-n<0~\text{or}~2l-n>k, \\
  a_{2l-n}, & \text{otherwise}.
  \end{cases}
\end{align}
The stability conditions can be satisfied when all the determinants of the matrices $ \theta_{k} $ are positive~\cite{SDeJesus1987PRA}. Through careful analysis, we find only $\theta_{5}$ and $\theta_{6}$ are nontrivial. As shown in Figs.\,\ref{stability}(b) and \ref{stability}(c), we numerically plot these functions in a logarithmic form, i.e.,
\begin{align}
\Theta_{k}
= \begin{cases}
  \ln\theta_{k}, & \theta_{k}>0, \\
  -\ln\left|\theta_{k}\right|, & \theta_{k}<0.
  \end{cases}
\end{align}
Note that $ \theta_{k} $ and $ \Theta_{k} $ maintain the same sign within the parameters used in the main text, thus the contour line in Fig.\,\ref{stability}(c) clearly determines the boundary between the stability and instability regions. In this case, we can confirm that these experimentally feasible parameters keep this optomechanical system in a stable zone.

\section{The role of thermal effects and quality factors}

Thermal noises can destroy fragile quantum correlations in practical devices. Thus protecting quantum resources from environmental thermal perturbations is essential for achieving quantum nonreciprocity. To see the influence of the thermal effect on nonreciprocal optomechanical entanglement, we plot the logarithmic negativity $ E_{\mathcal{N}} $ with respect to the environment temperature $ \textrm{T} $ in Figs.\,\ref{thermaleffect}(a)\,-\,\ref{thermaleffect}(d). We find that nonreciprocal optomechanical entanglement in a chosen direction can exist at higher temperature compared to the case of a stationary system. This means that, by spinning the resonator, optomechanical entanglement can be more robust against thermal noises. In addition, by defining the difference of COM entanglement for the opposite driving directions: $ \Delta E_{\mathcal{N}}\equiv E_{\mathcal{N},\circlearrowleft}-E_{\mathcal{N},\circlearrowright} $, we demonstrate the dependence of quantum nonreciprocity on environmental temperature in Figs.\,\ref{thermaleffect}(e) and \ref{thermaleffect}(f). Clearly, the condition $ \Delta E_{\mathcal{N}}\neq 0 $ can be satisfied even at $ \textrm{T}\sim 600\,\textrm{mK} $. Thereby, quantum nonreciprocity provides a new strategy to protect quantum entanglement in a noisy environment, i.e., enhancing the entanglement quality in a chosen (wanted) direction at the price of losing its quality in the other (unwanted) direction. This new possibility, as far as we know, has not been revealed in all previous works.
\begin{figure}[t]
\centering
\includegraphics[width=0.95\textwidth]{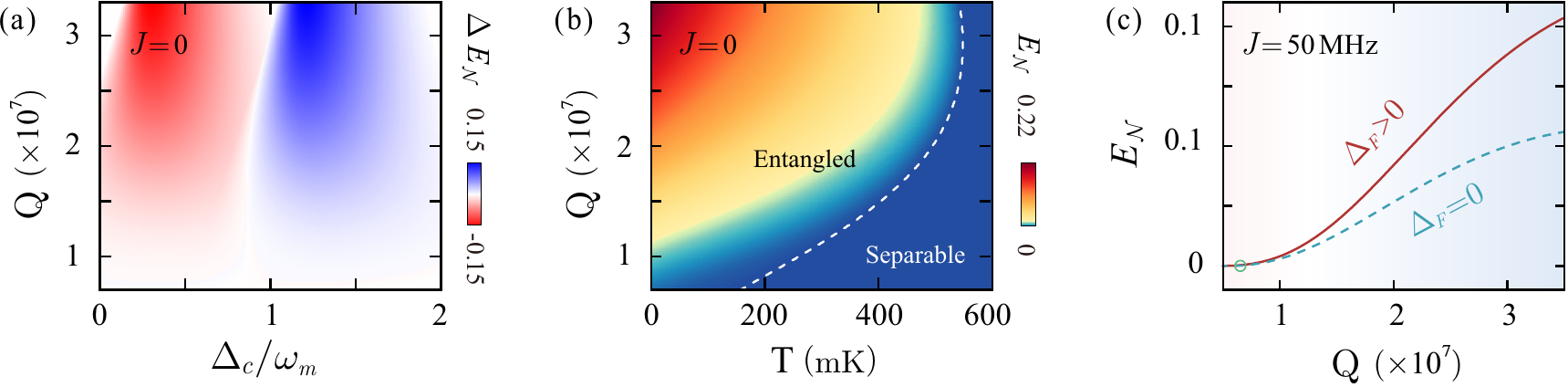}
\caption{\label{Qfactor}The influence of $ Q $-factor on nonreciprocal optomechanical entanglement. (a) The COM entanglement difference, $ \Delta E_{\mathcal{N}} $, versus the scaled optical detuning $ \Delta_{c}/\omega_{m} $ and the quality factor $Q$ at temperature $ \textrm{T}=130\,\textrm{mK} $. (b) Density plot of the logarithmic negativity $ E_{\mathcal{N}} $ versus the environment temperature $ \textrm{T} $ and the quality factor $Q$ at $ \Delta_{c}/\omega_{m}=1.2 $. (c) The logarithmic negativity $ E_{\mathcal{N}} $ as a function of the quality factor $Q$ at $ \Delta_{c}/\omega_{m}=1.6 $. The rotation speed is chosen as $ \Omega=8\,\textrm{kHz} $ in (a)-(c). Other parameters are listed in Table~\ref{tab:parameters}.}
\end{figure}

In addition, the $Q$-factor of the resonator also affects the realization of quantum nonreciprocity. In case of $ J=0 $, as shown in Fig.\,\ref{Qfactor}(a), it is found that quantum nonreciprocity can persist for $ Q\geq 10^{7} $ at $ \textrm{T}=130\,\textrm{mK} $ but its degree is lowered by decreasing the values of $Q$. Fig.\,\ref{Qfactor}(b) shows that both lower $Q$-factor and higher temperature are harmful for the robustness of COM entanglement. In addition, it is seen that COM entanglement even completely vanishes at high temperatures regardless of $Q$-factors. Moreover, in the presence of backscattering, e.g., $ J=50\,\textrm{MHz} $, it is found that for the same values of $Q$ and $ \textrm{T} $, by spinning the resonator, nonreciprocal entanglement can always be better than that in a nonspinning device [see Fig.\,\ref{Qfactor}(c)]. These results indicate that considerable entanglement revival can be achieved with high $ Q $-factor spinning resonators.

\end{document}